\documentclass[10pt]{article}
\usepackage[OE]{express}
\usepackage{graphicx}
\usepackage[caption=false]{subfig}
\usepackage{epstopdf}
\usepackage{float}

\begin{document}
	
	\title{Multi-point Dynamic Strain Sensing using External Phase Modulation-based Brillouin Optical Correlation Domain Analysis}
	
	\author{Bhargav Somepalli, Deepa Venkitesh, and Balaji Srinivasan\authormark{*}}
	
	\address{Department of Electrical Engineering, Indian Institute of Technology Madras, Chennai 600036, India}
	
	\email{\authormark{*}balajis@ee.iitm.ac.in}
	
	
	\begin{abstract}
		We report a novel technique to detect dynamic strain variations simultaneously at multiple locations. Our technique is based on Brillouin optical correlation domain analysis implemented through external phase modulation to generate multiple independently-accessible correlation peaks within the sensing fiber. Simulations are carried out to demonstrate the precise determination of Brillouin frequency shift (BFS) from multiple locations independently, and the corresponding results are validated through controlled experiments. Two correlation peaks are generated within a 1 km long fiber and their independent tunability is verified experimentally by mapping the spatial profile of the two correlations. We demonstrate the detection of dynamic strain variations up to 50 Hz (limited only by our test setup) at two locations, each with a spatial resolution of 6 m over 1.1 km long fiber.
	\end{abstract}
	
	

	
	\section{Introduction}
	Stimulated Brillouin scattering (SBS) based sensors have been extensively studied and implemented in the past few decades for distributed measurement of strain and/or temperature \cite{Horiguchi_strain,Kurashima_temperature,Thevenaz_strain_temp,Long_range_sensing,Bao_SBS_review}. Specifically, Brillouin optical correlation domain analysis (BOCDA) \cite{Hotate_2000,zadok2012random} has been widely used for dynamic strain sensing applications requiring sub-meter spatial resolution \cite{BOCDA_dynamic} such as load monitoring of aircrafts\cite{Aircraft_SHM}. In BOCDA, a  frequency-modulated \cite{hotate2002distributed} or phase-modulated \cite{antman2012localized} narrow linewidth laser source is split into two to generate the pump and probe lightwaves. These two lightwaves are counter propagated in the fiber under test (FUT) thereby localizing the SBS process through correlation peaks formed at specific periodic locations.
	
	Conventional frequency-modulated BOCDA sensors employ direct modulation of a narrow linewidth laser source such as a distributed feedback (DFB) laser to generate frequency modulated (FM) pump and probe\cite{song2006optimization}. This technique is useful to generate only one independent correlation peak within the FUT, thus limiting the measurements to a single location at any given instance. Distributed sensing is achieved by sweeping the same correlation peak along the FUT by varying the modulation frequency ($f_m$) \cite{tanaka2002application}. However, several SHM applications require faster measurements ($\sim$100 Hz) as well as more number of sensing points. This can be addressed by enabling simultaneous measurements at multiple locations. Although several configurations have been demonstrated to monitor multiple locations such as random access BOCDA \cite{Random_access} and temporal gating BOCDA \cite{Temporal_gating,elooz2014high}, multiple locations are monitored sequentially in all such configurations, thereby limiting the measurement speed. We recently proposed external phase modulation-based BOCDA as a technique to monitor multiple independent locations simultaneously \cite{Acp_2017}. We also experimentally demonstrated the generation and control of multiple correlation peaks within the FUT \cite{Acp_2017}.
	
	In this paper, we demonstrate the use of external phase modulation-based BOCDA to measure dynamic strain events at different locations in the sensing fiber. To the best of our knowledge, this is the first ever experimental demonstration of independent control of multiple correlation peaks in BOCDA which enables simultaneous multi-point dynamic strain sensing. We present simulations on extracting the Brillouin gain spectrum (BGS) from multiple correlation peak locations independently, which are validated through controlled experiments. Static and dynamic strain variations are applied to the FUT and the corresponding changes in the Brillouin frequency shift (BFS) are detected using phase modulation-based BOCDA. The paper is organized as follows. The principle of phase modulation-based BOCDA is discussed in Section 2. This is followed by detailed simulations to demonstrate the extraction of strain variations from multiple correlation peak locations in the third section. The experimental details and the results are discussed in the subsequent sections. This is followed by a discussion of the limitations in scaling the technique to multiple locations and conclusions from this work.
	
	\section{Principle of Phase Modulation-based BOCDA}
	The BOCDA system typically uses direct modulation of a narrow linewidth source with sinusoidal signal to achieve frequency modulated pump and probe, which in turn results in periodic correlations with a correlation at the center of the fiber - referred to as the zeroth order correlation peak. The separation between the adjacent correlations ($d$) is given by \cite{Hotate_review}
	\begin{equation}
	d = \frac{c}{2 n f_m} \sim \frac{10^8}{f_m} \label{d},
	\end{equation}
	where $c$ is the speed of light in vacuum, $n$ is the effective index of the fundamental mode in fiber and $f_m$ represents the modulation frequency. The location of the correlation peaks (except for the zeroth order peak) can be tuned by changing the modulation frequency $f_m$. The full-width at half maximum of each correlation ($\Delta z$) is given by \cite{Hotate_review}
	\begin{equation}
	\Delta z = \frac{c \Delta \nu_B}{2 \pi n f_m \Delta f} \label{dz},
	\end{equation}
	where $\Delta \nu_B$ is the Brillouin gain bandwidth ($\sim$ 30 MHz) and $\Delta f$ is the frequency deviation. Thus, for a given fiber, the location of sensing is solely determined by the modulation frequency, while the spatial resolution (which depends on the width of the correlation) is additionally influenced by the frequency deviation. In order to have unambiguous and reliable sensing in a given length of FUT, the modulation frequency is chosen such that only one correlation peak exists within the FUT thereby monitoring only one location. One of the two lightwaves - pump or probe, is delayed relative to the other such that the correlation peak generated within the FUT corresponds to non-zeroth interaction and hence can be tuned across the FUT for distributed sensing. In order to monitor multiple locations simultaneously, multiple correlation peaks have to be generated within the FUT. This would require optical modulation with signals that have different $f_m$ values, each of which would uniquely determine a sensing location. Additionally, each of these correlation peaks have to be tuned independently in order to monitor the user-specified locations. Direct modulation of laser is not amicable to the above constraints, the details of which are discussed in the Appendix.
	
	In this work, we generate the requisite sinusoidal FM signals with unique sets of $f_m$ and $\Delta f$ in the electrical domain using an arbitrary waveform generator and embed these features in the optical domain through an external phase modulator (PM).
	
	\begin{figure}[htbp]
		\centering\subfloat[]{\includegraphics[width=3.6cm]{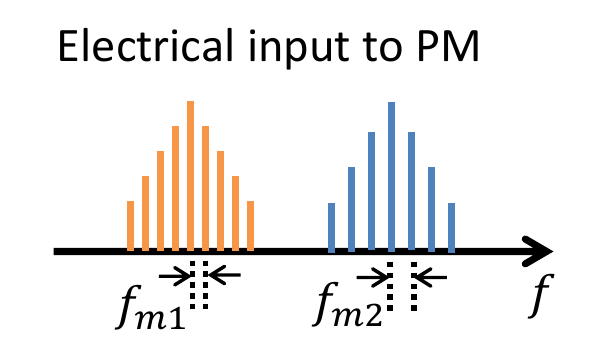}}
		\subfloat[]{\includegraphics[width=7.1cm]{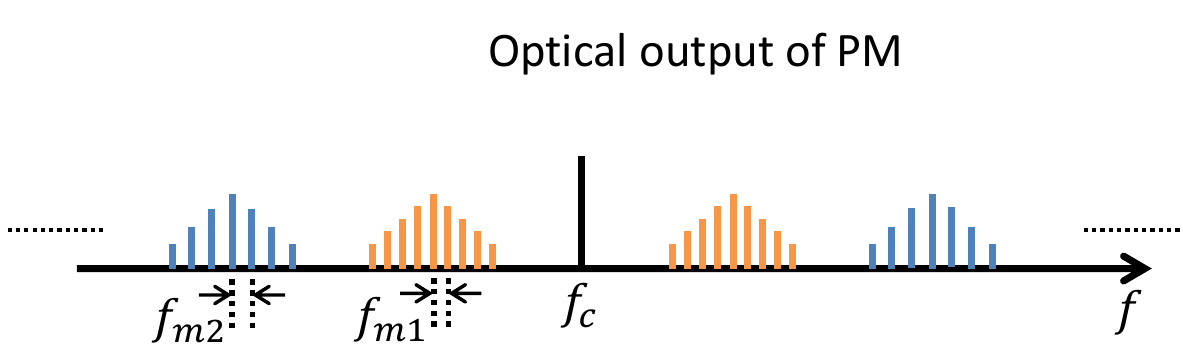}}\\
		\centering\subfloat[]{\includegraphics[width=11cm]{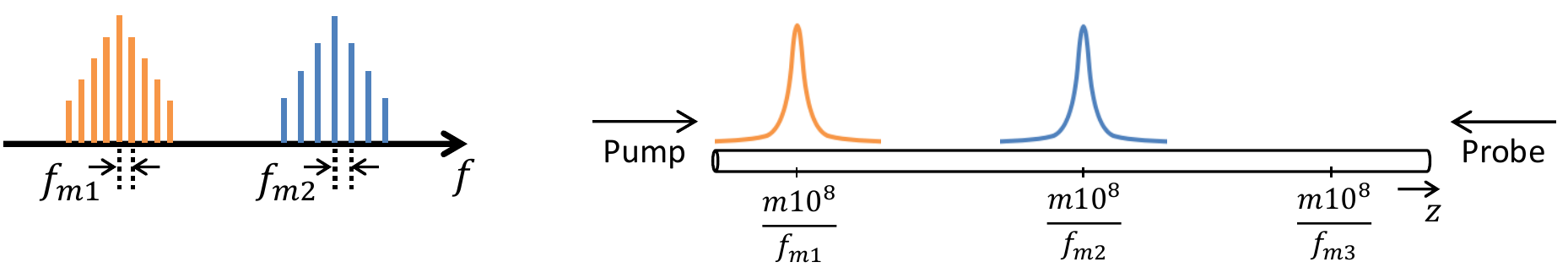}}\\
		\centering\subfloat[]{\includegraphics[width=11cm]{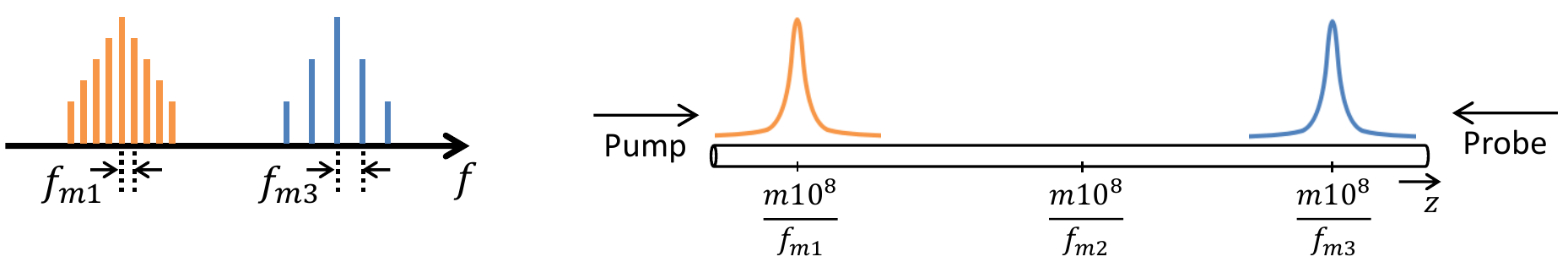}}
		\caption{\label{fig:FM_generation}Mechanism of generating multiple FM signals using external phase modulation. (a) Typical frequency spectra of an electrical signal with FM modulation at different center frequencies and (b) corresponding output spectrum of optical phase modulator. (c-d) Independent tunability of correlation peaks by modifying the modulation frequency of one of the FM signals from $f_{m2}$ to $f_{m3}$.}
	\end{figure}
	
	An optical phase modulator when driven by an electrical signal generates multiple sidebands with a spectral content similar to that of the driving electrical signal. The typical structure of the  frequency spectrum of the sinusoidal FM drive signal and that at the output of the phase modulator are shown in Figs. \ref{fig:FM_generation}(a) and \ref{fig:FM_generation}(b) respectively. The drive signal is comprised of multiple FM signals at distinct center frequencies and with different $f_m$ frequencies in the electrical domain as is shown in Fig. \ref{fig:FM_generation}(a). The frequency deviation $\Delta f$ decides the strength of the side bands for each FM set. If the FM sets are within the bandwidth of the phase modulator, the optical output of the same (shown in Fig. \ref{fig:FM_generation}(b)) is expected to have optical carrier (at $f_c$) and multiple FM signals on both sides of the optical carrier with same $f_m$ frequencies as in the electrical domain. When such an optical signal is used to generate the pump and the probe in a Brillouin sensing experiment, each of these multiple FM signals generates a corresponding correlation peak whose location and width are determined by the respective $f_m$ and $\Delta f$ values. The location of each correlation peak can be tuned independent of the other by modifying the corresponding FM signal in the electrical domain as shown in Figs. \ref{fig:FM_generation}(c) and \ref{fig:FM_generation}(d).
	
	Driving the phase modulator with two FM signals with modulation frequencies $f_{m1}$ and $f_{m2}$ (as shown in Fig. \ref{fig:FM_generation}(c)) generates two correlation peaks at locations given by Eq. (\ref{d}). When the modulation frequency of one of the FM signals is modified from $f_{m2}$ to $f_{m3}$, the location of the corresponding correlation peak alone can be changed as shown in Fig. \ref{fig:FM_generation}(d). Thus by driving the phase modulator using multiple FM signals and with a careful choice of $f_m$ values, multiple independently-addressable correlation peaks can be generated at specific locations across the sensing fiber, thereby enabling the ability to monitor multiple locations simultaneously. Even though the illustrations shown in Fig. \ref{fig:FM_generation} indicate distinct center frequencies for different FM signals, the difference between the center frequencies do not influence the measurement result; they could in fact be identical in an experiment.
	
	\section{Simulation results}
	
	One of the key challenges in monitoring multiple locations simultaneously is the extraction of the strain information from multiple sensing locations without significant cross-talk. Simulations are performed to verify the dependence of BGS from each correlation peak with that of the other correlation peaks in the FUT. We extend the methodology followed in \cite{Iop_mst} to simulate the SBS interaction over 1 km long fiber and estimate the amplified probe when the two lightwaves - pump and probe are modulated with multiple sinusoidal FM signals. Under undepleted pump approximation, SBS interaction between pump and probe is modeled using the steady-state propagation equations \cite{Agrawal}. The amplified probe power is computed using the pump power and SBS gain which depends on the local BFS and the instantaneous frequency offset between pump and probe \cite{Boyd}.
	
\begin{figure}[htbp]
		\centering\subfloat[]{\includegraphics[width=6cm]{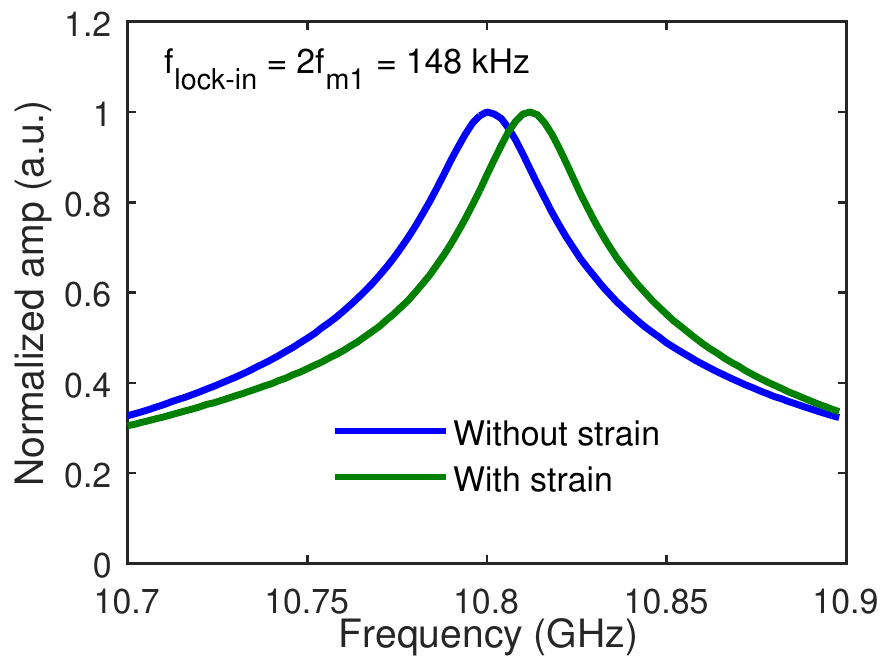}}
		\subfloat[]{\includegraphics[width=6cm]{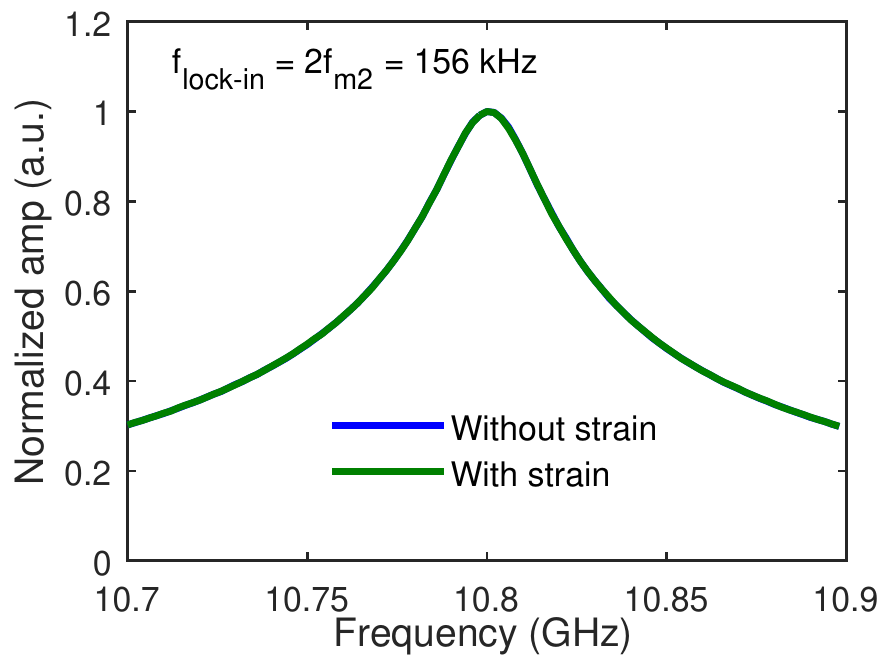}}
		\caption{\label{fig:Strain_BGS_simu}Simulated BGS traces at correlation peak locations corresponding to (a) $f_{m1}$ = 74 kHz and (b) $f_{m2}$ = 78 kHz. Strain is simulated at the location corresponding to the correlation peak at $f_{m1}$ = 74 kHz. BGS traces are obtained through lock-in detection corresponding to 2$f_m$ frequencies.}
	\end{figure}

	The time step size considered is 5 ns which corresponds to space step size of 1 m. The BFS of the fiber is considered as 10.800 GHz. The pump and probe are considered to be modulated with two sinusoidal FM signals centered at 6 GHz with $f_m$ frequencies 74 kHz and 78 kHz and $\Delta f$ of 2 GHz each. The probe is delayed by 70 $\mu$s relative to the pump. This generates two correlation peaks at 450 m and 800 m as per Eq. (\ref{d}). The spatial resolution, given by the width of correlation (Eq. (\ref{dz})), is nearly 6 m each. The frequency offset between pump and probe is varied from 10.700 GHz to 10.900 GHz. A strain perturbation equivalent to an increase in BFS of 10 MHz is simulated at the correlation peak location which corresponds to an $f_m$ frequency of 74 kHz. The BGS traces are obtained by simulating lock-in detection \cite{Lock_in} at the corresponding 2$f_m$ frequencies sequentially. The BGS at the two correlation peak locations obtained through simulations in the presence and absence of strain are shown in Fig. \ref{fig:Strain_BGS_simu}.
	
	In the absence of strain, the BFS at the two locations is 10.800 GHz. In the presence of strain, the peak of the BGS at the correlation peak location corresponding to a modulation frequency of 74 kHz is shifted to 10.81 GHz while the other peak corresponding to a modulation frequency of 78 kHz has not shifted. This conveys that the BGS of each correlation peak location is independent on the BGS of the other correlation peak locations. These results are validated through experiments as described in the following sections.
	
	\section{Experimental details}
	
\begin{figure}[htbp]
		\centering\includegraphics[width=13cm]{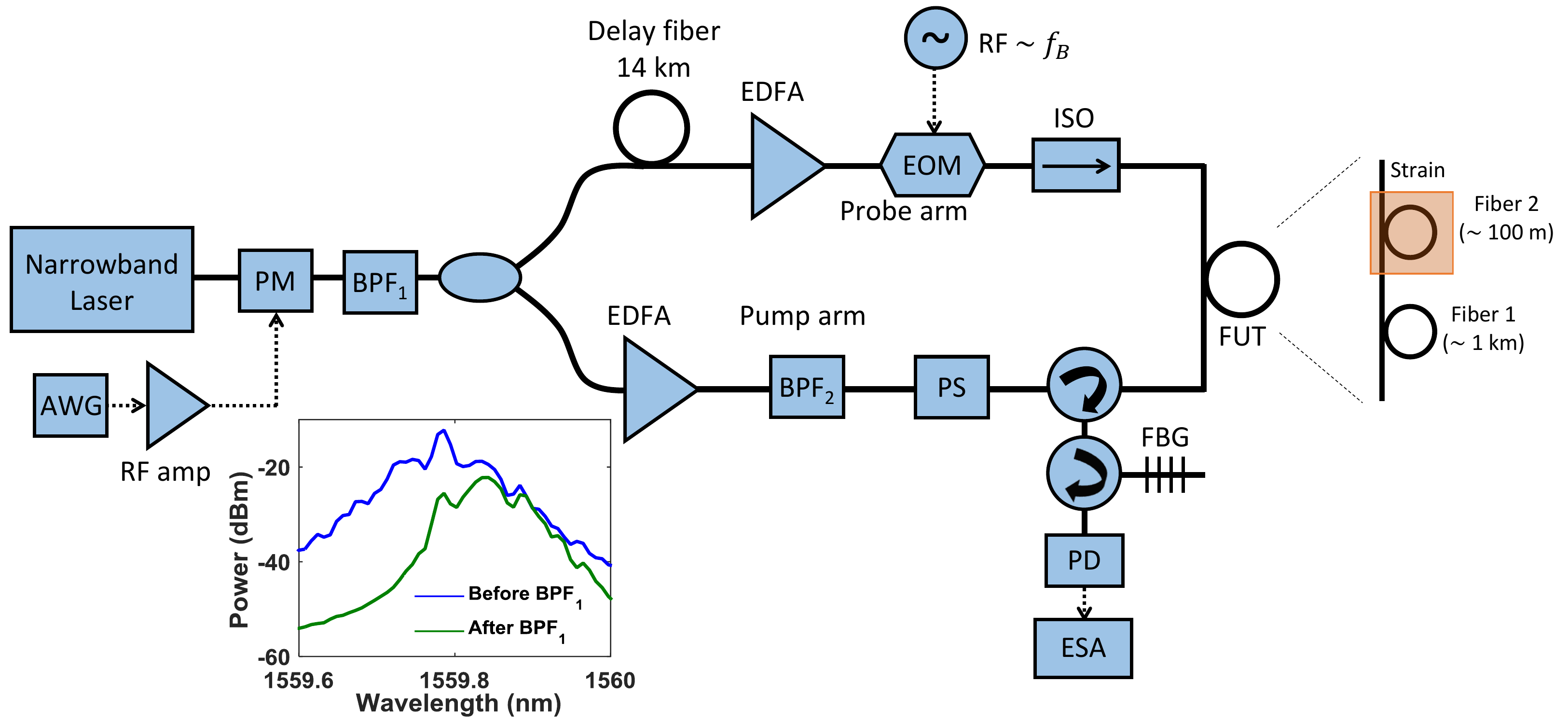}
		\caption{\label{fig:Expt_setup}Schematic of the experimental setup and the fiber under test. The optical spectra before and after the BPF$_1$ are shown in the inset; PM: Phase modulator; AWG: Arbitrary waveform generator; BPF: Bandpass filter; EDFA: Erbium-doped fiber amplifier; EOM: Electro-optic modulator; ISO: Optical isolator; PS: Polarization scrambler; FBG: Fiber Bragg grating; PD: Photo detector; ESA: Electrical spectrum analyzer.}
	\end{figure}

	A schematic diagram of the experimental setup used to monitor multiple locations using phase modulation-based BOCDA is shown in Fig. \ref{fig:Expt_setup}.
	A narrowband laser (Coherent Solutions - linewidth 25 kHz) at a wavelength of $\sim$1560 nm is used as a light source. The output of the laser is modulated using an external phase modulator (Photline - MPZ-LN-10), which is driven by the sum of two sinusoidal FM signals generated from an arbitrary waveform generator (AWG - Keysight M8195A). The two FM signals are centered at 6 GHz with a frequency deviation ($\Delta f$) of 2 GHz each. The modulation frequencies of the two FM signals are varied between 71 kHz and 80 kHz, which corresponds to a measurement range and spatial resolution of about 1.3 km and 6 m respectively according to Eqs. (\ref{d}) and (\ref{dz}). The output of the phase modulator is filtered using a bandpass filter (BPF$_1$ - Finisar WaveShaper 1000S) to extract the frequency modulated optical signals (shown in inset of Fig. \ref{fig:Expt_setup}) which are subsequently split into pump and probe waves. The pump lightwave after amplification is launched from one end of the FUT consisting of a 1 km long fiber (Fiber 1) followed by a 100 m long fiber (Fiber 2). The probe lightwave on the other arm is passed through 14 km long delay fiber so that the correlation peak generated within the FUT corresponds to non-zeroth order interaction. The delayed probe is amplified, frequency shifted by the Brillouin frequency ($f_B$) using an electro-optic modulator (EOM) in carrier suppressed configuration and is launched from the other end of the FUT.
	
	The frequency modulated pump and probe interact in the FUT and generate multiple correlation peaks at locations determined by the carefully chosen $f_m$ frequencies. The amplified probe is filtered using a fiber Bragg grating to extract the Brillouin Stokes component and is detected using a 45 MHz photo receiver. Lock-in detection at 2$f_m$ frequency is performed using an electrical spectrum analyzer (ESA - R$\&$S FSV30) in zero-span mode.
	
	\section{Results and Discussion}
	\subsection{Spatial Mapping of Multiple Correlation Peaks}
\begin{figure}[htbp]
		\centering{\includegraphics[width=6cm]{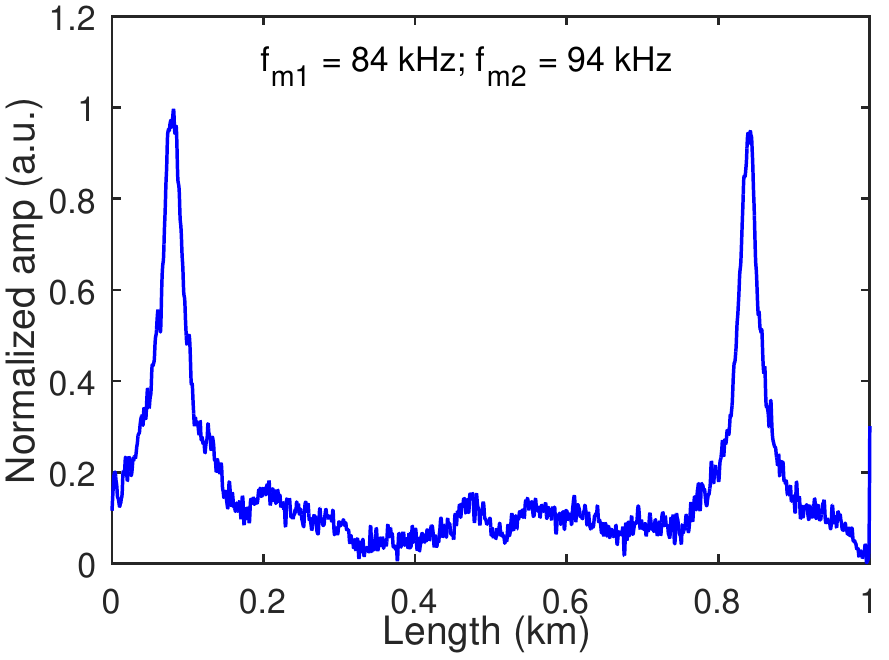}}
		\caption{\label{fig:Two_corr_peak_profile}Amplified probe trace obtained by pulsing the pump showing two correlation peaks when phase modulator is driven with two FM signals \cite{Acp_2017}.}
	\end{figure}

	As mentioned in the Introduction, one of the key objectives of this work is to generate multiple correlation peaks in the FUT simultaneously. In order to verify the generation and independent control of multiple correlation peaks, initial tests are carried out with only Fiber 1 ($\sim$1 km) as the FUT. We follow the same procedure as in \cite{Iop_mst} in order to spatially map the correlation profiles to confirm their actual locations. In order to enable spatial mapping, we modulate the pump with a narrow pulse train (50 ns pulsewidth, 11 $\mu$s period) using an EOM \cite{Iop_mst} and observe the amplified probe on an oscilloscope. The phase modulator is driven with two sinusoidal FM signals with modulation frequencies ($f_m$) of 84 kHz and 94 kHz respectively and $\Delta f$ of 500 MHz each, which corresponds to a measurement range and spatial resolution of 1.1 km and 21 m respectively according to Eq. (\ref{d}) and (\ref{dz}). The amplified probe trace observed on an oscilloscope for 10.800 GHz frequency offset between pump and probe is shown in Fig. \ref{fig:Two_corr_peak_profile} \cite{Acp_2017}. Note that the time axis in the plot is translated to corresponding distances using the time of flight of pump.
	
\begin{figure}[htbp]
		\centering\subfloat[]{\includegraphics[width=6cm]{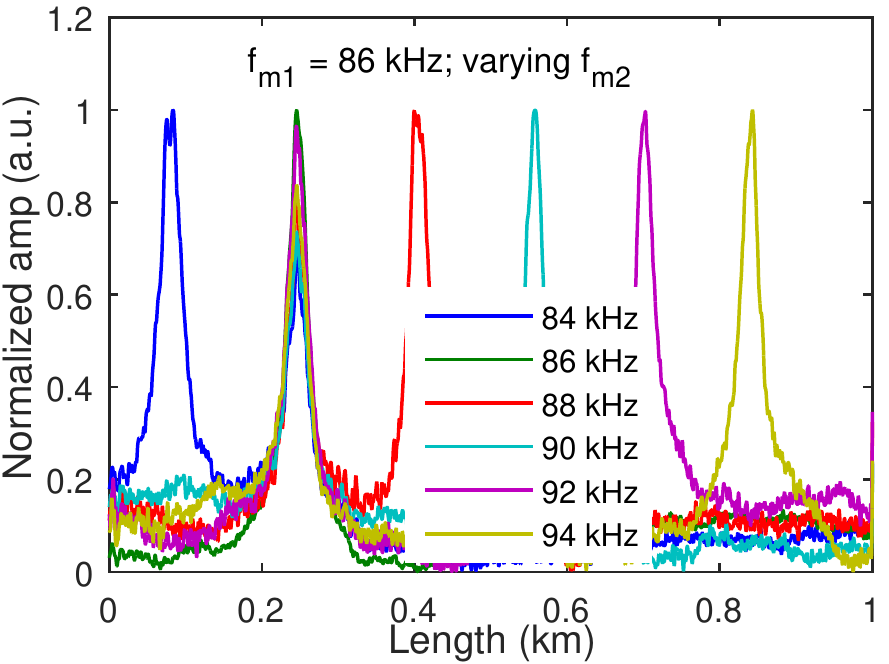}}
		\subfloat[]{\includegraphics[width=6cm]{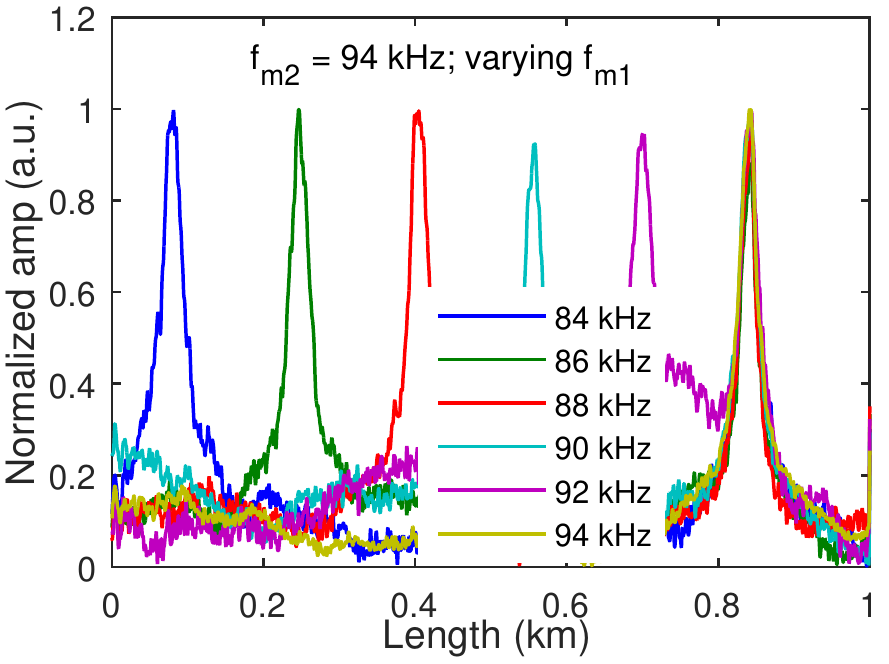}}
		\caption{\label{fig:Corr_profile_tuning}Traces showing the independent tunability of the two correlation peaks \cite{Acp_2017}.}
	\end{figure}

	The trace contains two distinct peaks indicating that two correlation peaks are generated due to the two FM signals at locations determined by the respective modulation frequencies ($f_{m1}$ and $f_{m2}$). The width of the correlation features are 40 m and 34 m respectively which have been verified independently through simulations. In order to demonstrate the independent tunability of the two correlation peaks, the modulation frequency ($f_{m2}$) of one of the FM signals is varied from 84 kHz to 94 kHz while that of the other is unchanged. The amplified probe traces obtained are shown in Fig. \ref{fig:Corr_profile_tuning}(a) \cite{Acp_2017}.
	
	The location of correlation peak corresponding to the varying modulation frequency $f_{m2}$ alone has changed while the one due to the fixed modulation frequency $f_{m1}$ remain unchanged. This has been also checked by keeping the modulation frequency $f_{m2}$ fixed and varying the other modulation frequency $f_{m1}$ as shown in Fig. \ref{fig:Corr_profile_tuning}(b). As expected, the width of each correlation feature is observed to change from 40 m to 34 m with change in $f_m$ from 84 kHz to 94 kHz. Both the results of Fig. \ref{fig:Corr_profile_tuning} clearly demonstrate that multiple correlation peaks generated through external phase modulation-based BOCDA can be tuned independently.
	
	\subsection{Sensing of static strain from multiple correlation peaks}
\begin{figure}[htbp]
		\centering{\includegraphics[width=6cm]{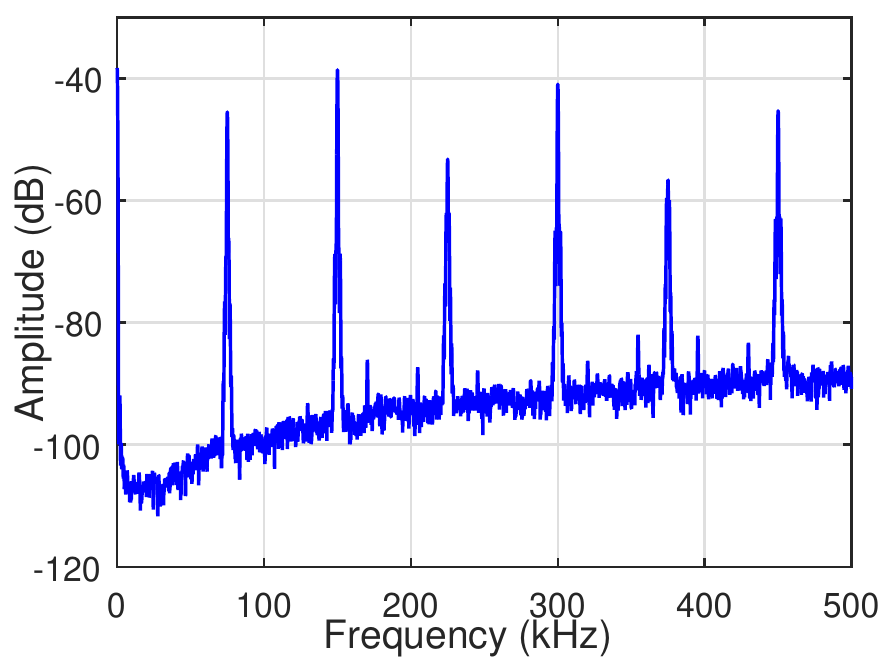}}
		\caption{\label{fig:75_spectrum}Spectrum recorded at the output of the photo detector when phase modulator is driven with one FM signal with $f_m$ of 75 kHz.}
	\end{figure}

	In order to detect static strain variations using phase modulation-based BOCDA, Fiber 2 ($\sim$100 m) is added to the FUT consisting of Fiber 1 ($\sim$1 km). FM signals with modulation frequencies between 70 kHz and 80 kHz are used to ensure that only one correlation peak is generated within the FUT due to each of the FM signals. Varying the modulation frequency from 71 kHz to 79 kHz sweeps the correlation peak across the Fiber 1 and with a modulation frequency closer to 80 kHz, the other correlation peak is localized within Fiber 2. In order to obtain the BGS and estimate the BFS along the FUT, the phase modulator is initially driven by one FM signal with $\Delta f$ of 2 GHz. The spectrum of the amplified probe after photo detection with a modulation frequency of 75 kHz is shown in Fig. \ref{fig:75_spectrum}.
	
\begin{figure}[htbp]
		\centering\subfloat[]{\includegraphics[width=6cm]{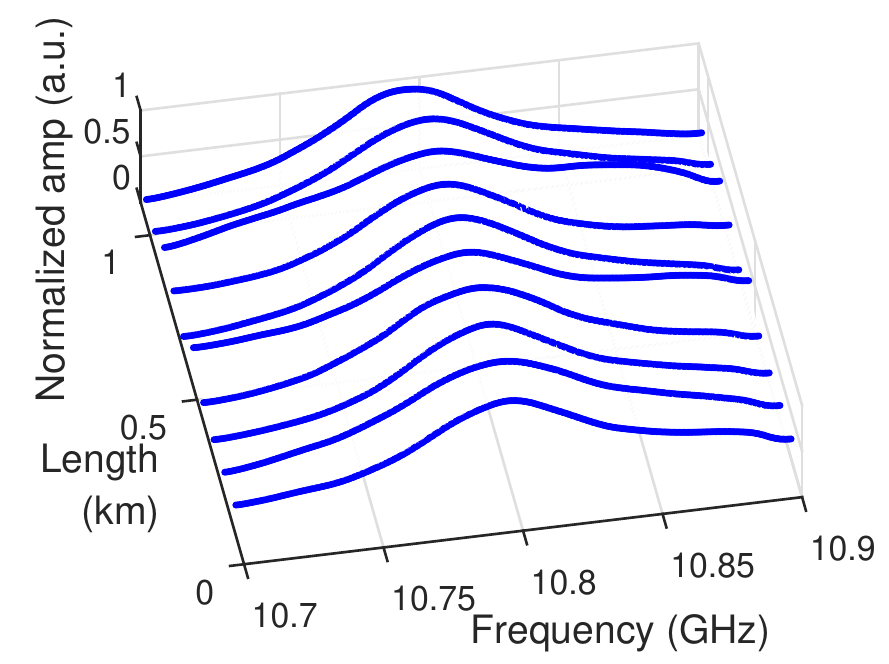}}
		\subfloat[]{\includegraphics[width=6cm]{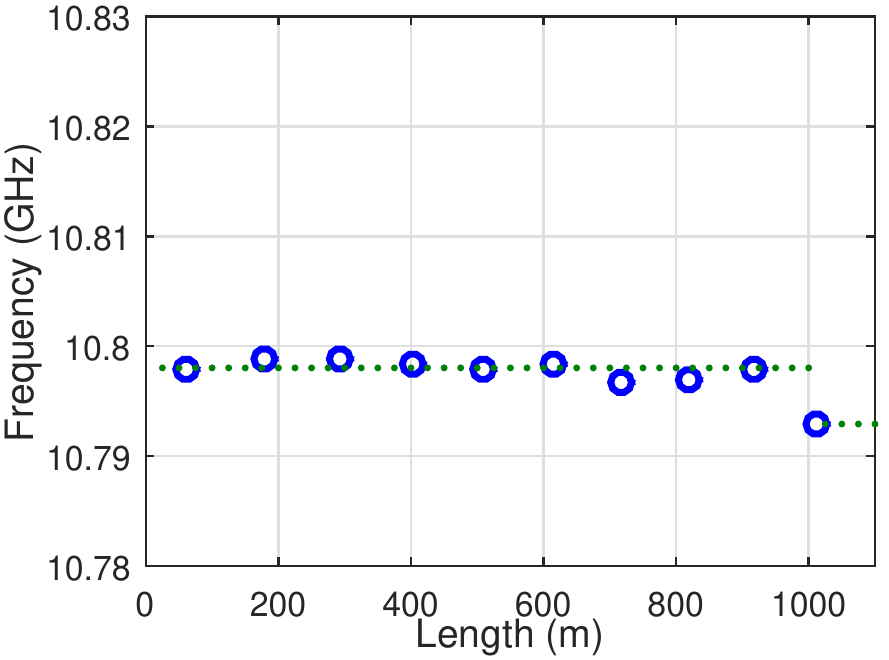}}
		\caption{\label{fig:Distributed_BGS}(a) BGS along 1.1 km long FUT obtained by varying $f_m$ from 71 kHz to 80 kHz and (b) the corresponding BFS as a function of the sensing fiber length.}
	\end{figure}

	The spectrum consists of distinct peaks at a frequency of $f_m$ and its harmonics. The frequency offset between pump and probe is varied from 10.701 GHz to 10.900 GHz in steps of 1 MHz and the corresponding BGS is captured using the ESA in zero-span mode locked to 2$f_m$ frequency. We experimentally verified the efficacy of choosing 2$f_m$ frequency for lock-in detection as opposed to other harmonics through independent experiments. The BGS is acquired for different $f_m$ values which sweeps the correlation peak along the 1.1 km long FUT. The BGS traces and the corresponding BFS obtained from peak detection at different locations are shown in Fig. \ref{fig:Distributed_BGS}.
	
\begin{figure}[htbp]
		\centering\subfloat[]{\includegraphics[width=6cm]{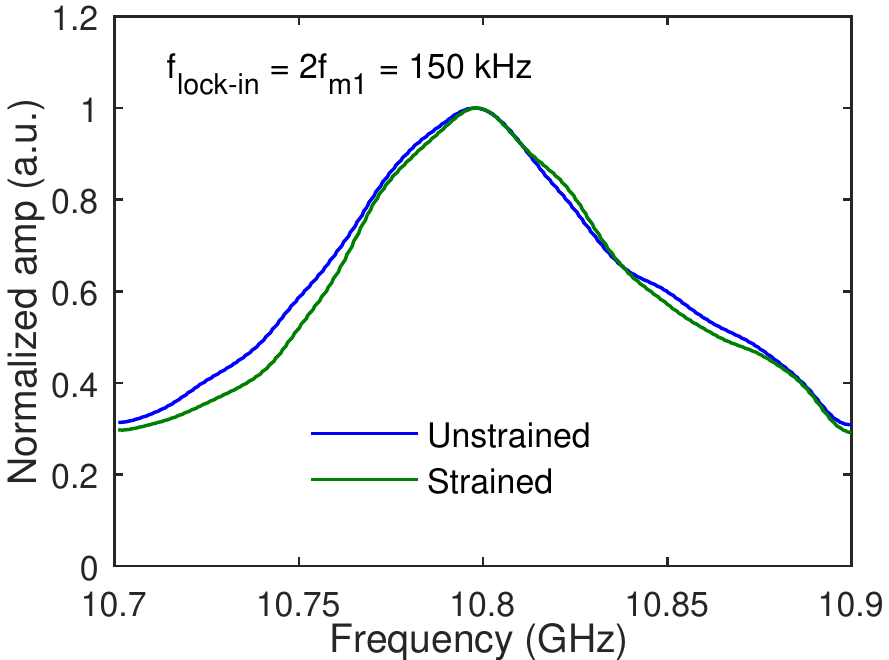}}
		\subfloat[]{\includegraphics[width=6cm]{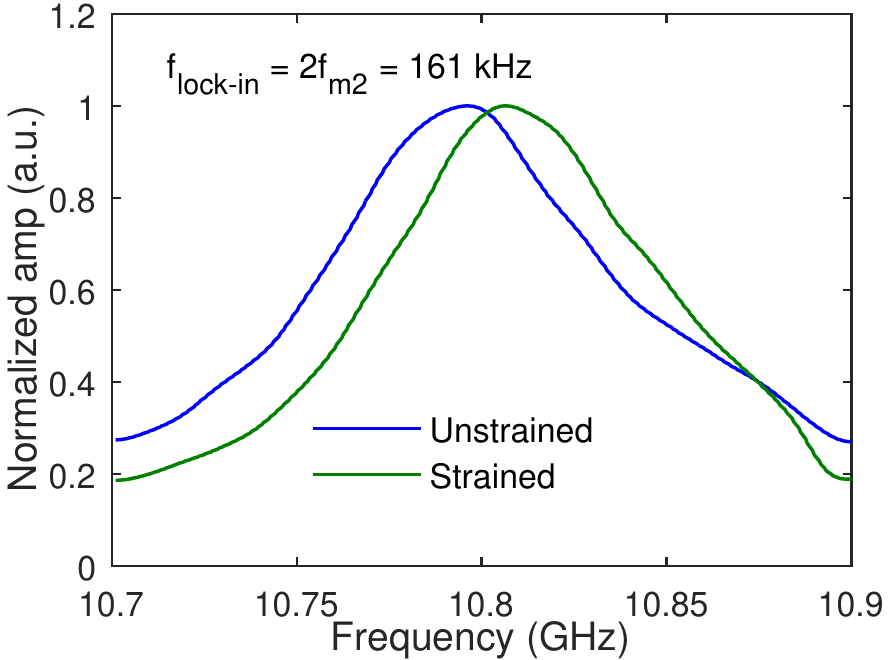}}
		\caption{\label{fig:75_80_static}BGS traces corresponding to the correlation peaks at (a) 510 m and (b) 1057 m generated using $f_{m1}$ = 75 kHz and $f_{m2}$ = 80.5 kHz respectively when strain is applied on Fiber 2.}
	\end{figure}

	The BFS of Fiber 1 is nearly 10.798 GHz while that of Fiber 2 is slightly lower ($\sim$10.793 GHz). The BFS of these fibers are measured independently through Brillouin optical time domain analysis measurements which are in good agreement with these values. We then proceed to drive the phase modulator with two sinusoidal FM signals centered at 6 GHz with $f_m$ frequencies of 75 kHz and 80.5 kHz such that correlation peaks are generated in the 1 km fiber and 100 m fiber respectively. Fiber 2 is wound across two posts mounted on translational stages and static strain is applied by moving one of the stages. The BGS traces obtained by locking ESA to the corresponding 2$f_m$ frequencies in zero-span mode are shown in Fig. \ref{fig:75_80_static}.
	
	Due to strain in Fiber 2, the BGS of the correlation peak location within Fiber 1 has not shifted while that in Fiber 2 has shifted by 10.1$\pm$0.8 MHz which corresponds to a strain perturbation of 202$\pm$16 $\mu\epsilon$. Such a strain perturbation on Fiber 2 is verified independently using a fiber Bragg grating wound across the two posts. These experiments validate our proposed technique of phase modulation-based BOCDA to measure strain at independent locations in the fiber; the BGS obtained from each correlation peak is dependent only on the strain applied at that location. Due to the presence of multiple frequency modulated signals in the system, there could be an increase in  beat noise when compared to a system with single correlation peak. However, the lock-in detection process still allows the precise extraction of BGS from multiple locations as shown in Figs. \ref{fig:75_80_static}(a) and \ref{fig:75_80_static}(b).
	
	\subsection{Sensing of dynamic strain from multiple correlation peaks}
\begin{figure}[htbp]
		\centering{\includegraphics[width=6cm]{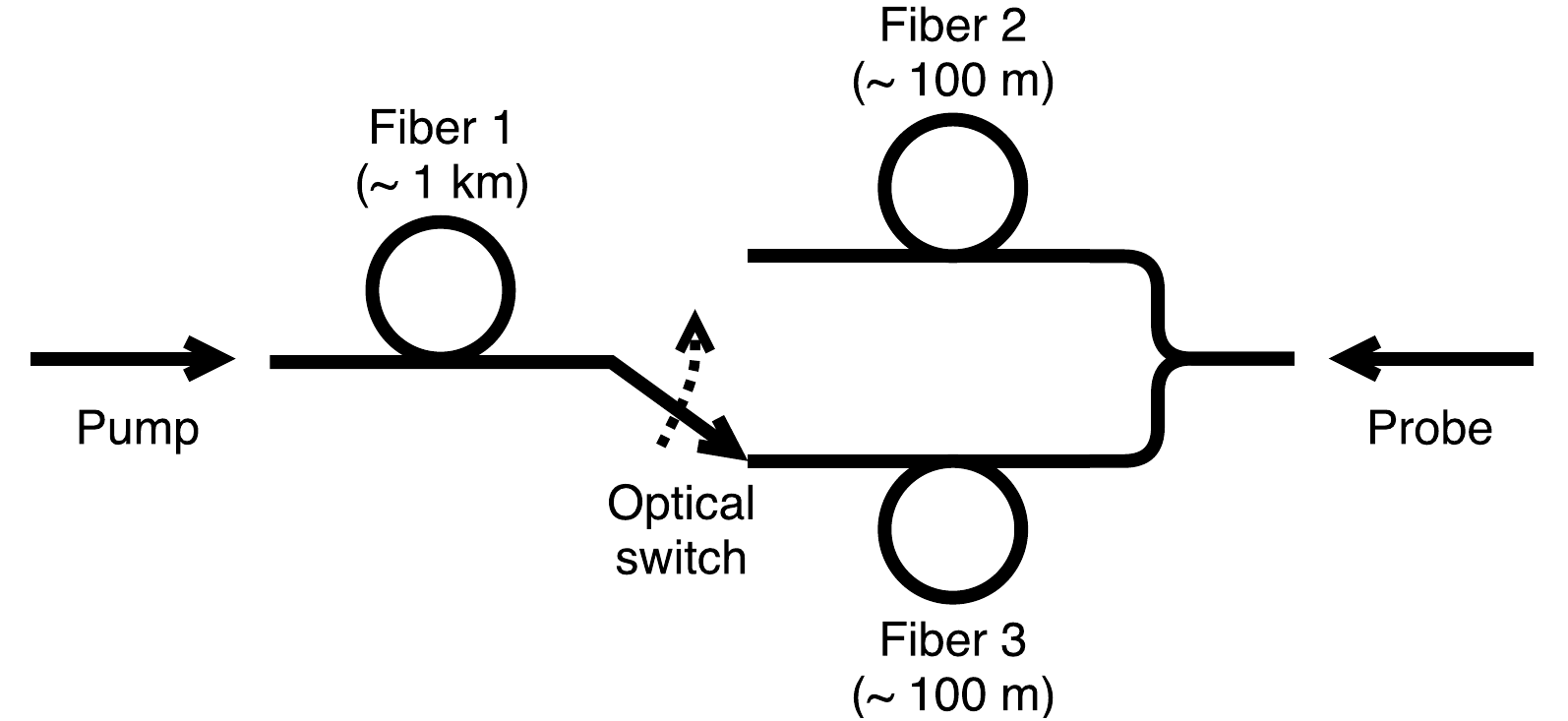}}
		\caption{\label{fig:FUT2}FUT to emulate dynamic strain over 100 m long fiber using an optical switch.}
	\end{figure}

	In order to demonstrate the capability of phase modulation-based BOCDA in detecting dynamic strain variations at multiple locations, dynamic strain is emulated in Fiber 2 by switching the optical path \cite{Optical_switch_ref} between two fibers whose Brillouin frequencies differ by 40 MHz approximately. The FUT is shown in Fig. \ref{fig:FUT2} where an optical switch (oeMarket - OSW-W1$\times$2; $T_{switch}$> 5 ms) is used to emulate dynamic strain in the 100 m long fiber.
	
\begin{figure}[htbp]
		\centering\subfloat[]{\includegraphics[width=6cm]{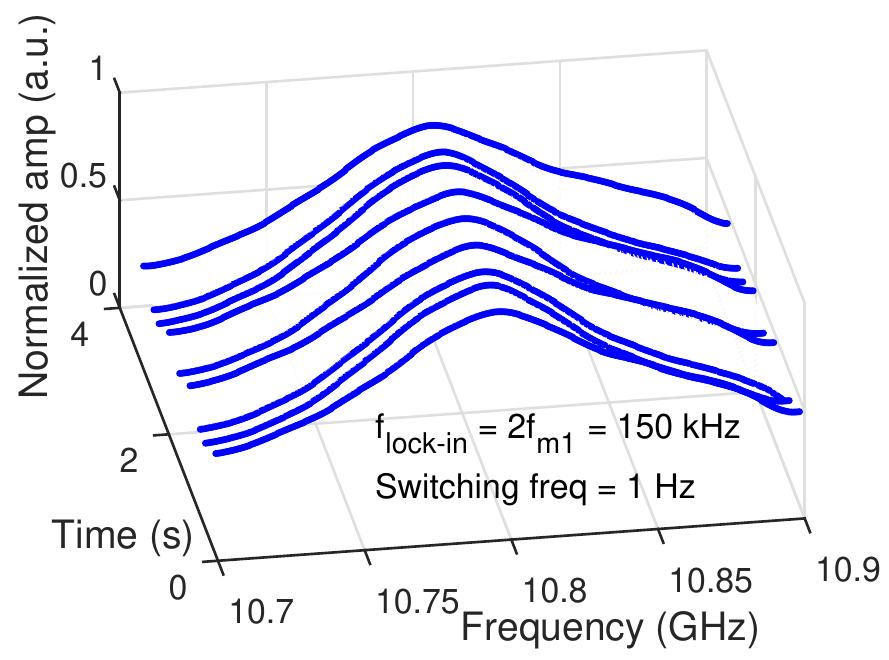}}
		\subfloat[]{\includegraphics[width=6cm]{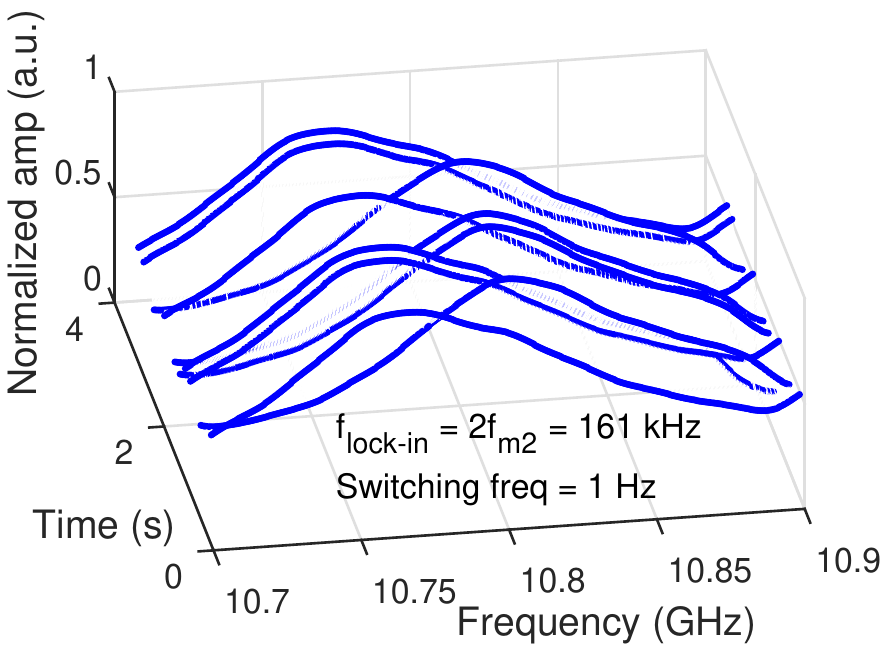}}\\
		\subfloat[]{\includegraphics[width=6cm]{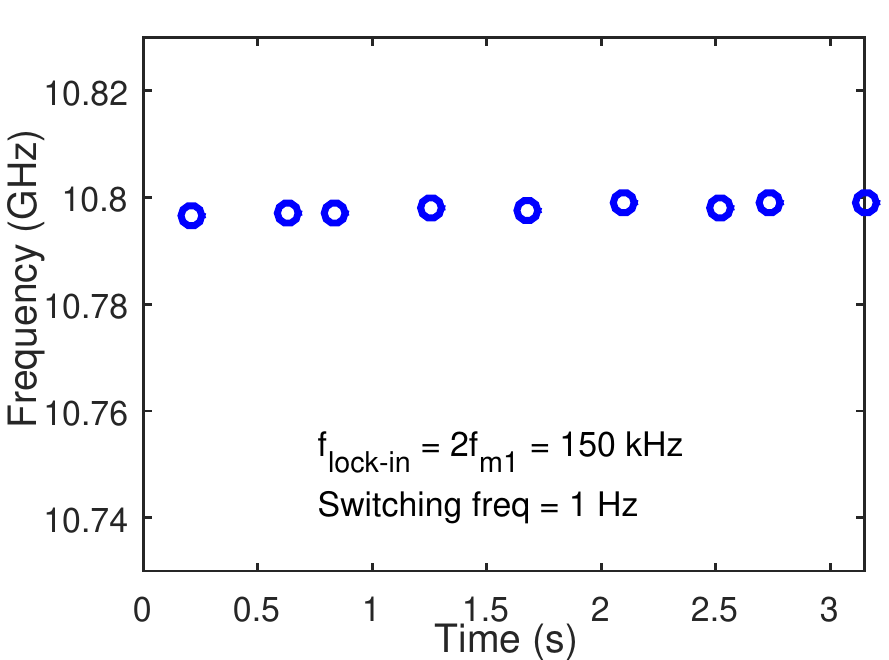}}
		\subfloat[]{\includegraphics[width=6cm]{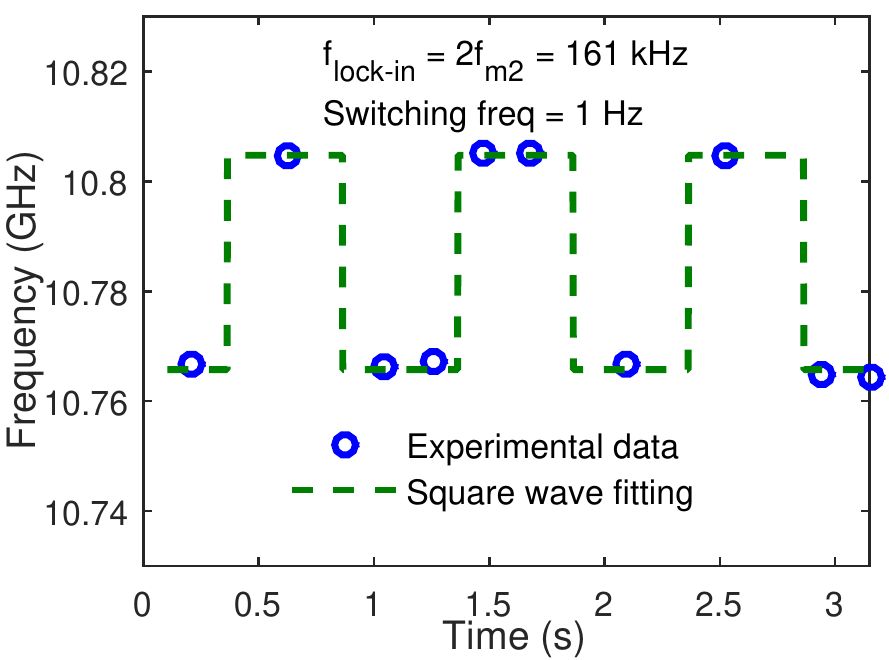}}
		\caption{\label{fig:75_80_BGS_500ms}BGS traces of the correlation peak location within (a) Fiber 1 ($f_{m1}$ = 75 kHz) and (b) Fiber 2 ($f_{m2}$ = 80.5 kHz) when Fiber 2 is subjected to dynamic strain with a switching frequency of 1 Hz. The corresponding BFS as a function of time are shown in (c) and (d). Step size of probe frequency scanning is 3 MHz.}
	\end{figure}

	Similar to the static strain measurements, two sinusoidal FM signals are used with $f_m$ frequencies of 75 kHz and 80.5 kHz and $\Delta f$ of 2 GHz each such that two correlation peaks are generated - one each in the 1 km fiber and the 100 m fiber. The optical path is switched between the two 100 m fibers, every 500 ms (switching frequency = 1 Hz). The frequency offset between pump and probe is varied from 10.701 GHz to 10.900 GHz in steps of 3 MHz; this process takes about 210 ms for each BGS measurement, limited only by the sweeping time of the probe frequency in our setup. Fig. \ref{fig:75_80_BGS_500ms} shows the BGS traces and the corresponding BFS as a function of time obtained through lock-in detection at the corresponding 2$f_m$ frequencies.
	
	The BGS of the correlation peak location within Fiber 1 is observed to remain unchanged as seen from Fig. \ref{fig:75_80_BGS_500ms}(a) and the corresponding BFS is almost constant as a function of time (Fig. \ref{fig:75_80_BGS_500ms}(c)). On the other hand, the BGS of the correlation peak generated within Fiber 2 is found to be switching periodically as seen from Fig. \ref{fig:75_80_BGS_500ms}(b) and the corresponding BFS is in good agreement with the square wave fitting with a switching frequency of 1 Hz (Fig. \ref{fig:75_80_BGS_500ms}(d)). The standard deviation in the estimated BFS is nearly 1.2 MHz. This demonstrates the multi-point dynamic strain sensing capability of the phase modulation-based BOCDA technique.
	
\begin{figure}[htbp]
		\centering\subfloat[]{\includegraphics[width=6cm]{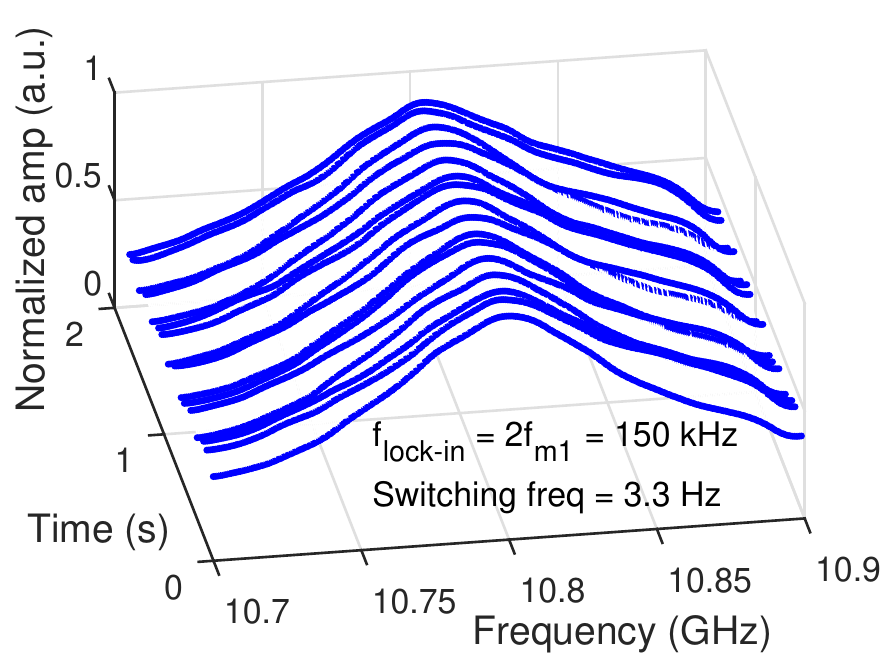}}
		\subfloat[]{\includegraphics[width=6cm]{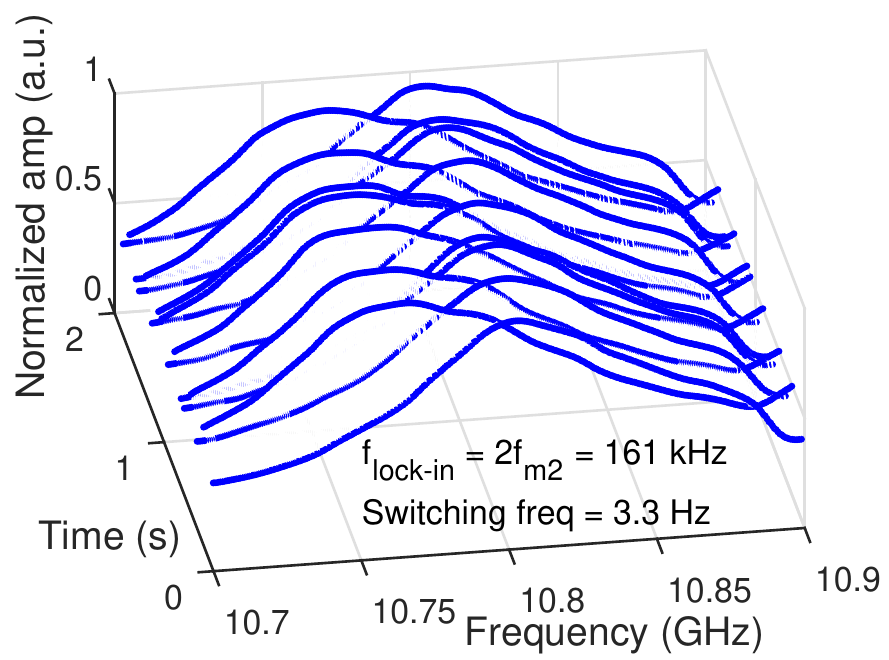}}\\
		\subfloat[]{\includegraphics[width=6cm]{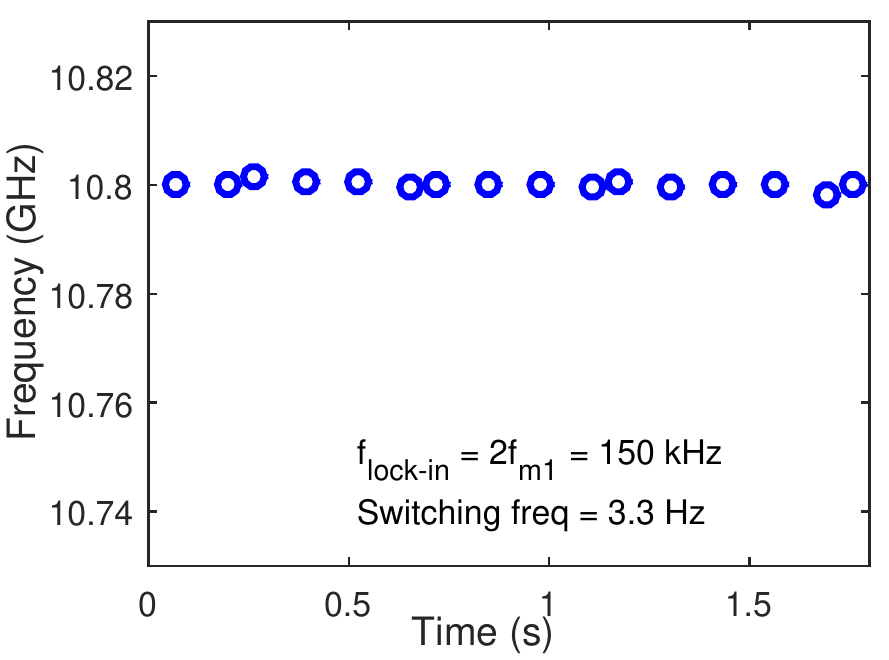}}
		\subfloat[]{\includegraphics[width=6cm]{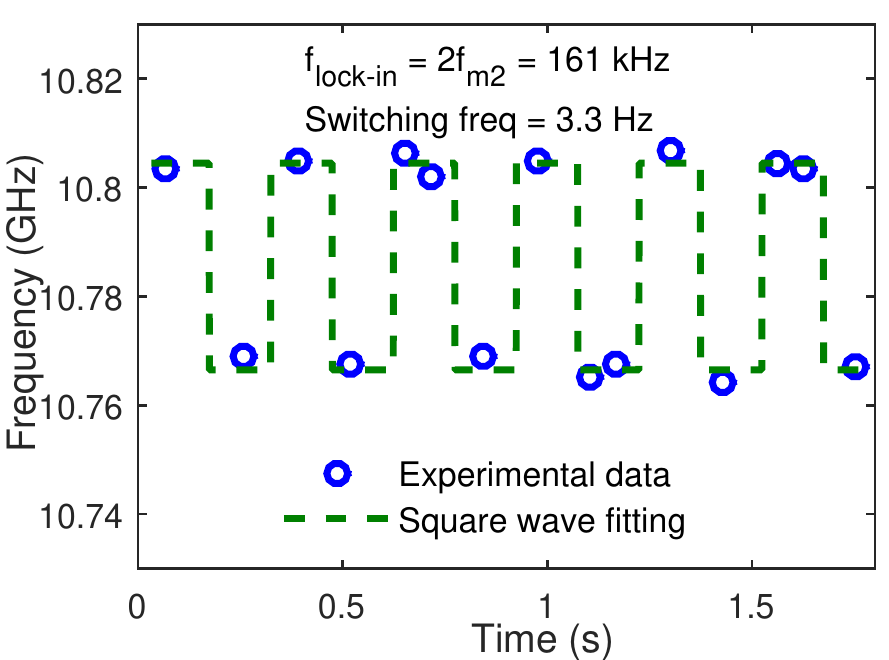}}
		\caption{\label{fig:75_80_BGS_150ms}BGS traces of the correlation peak location within (a) Fiber 1 ($f_{m1}$ = 75 kHz) and (b) Fiber 2 ($f_{m2}$ = 80.5 kHz) when Fiber 2 is subjected to dynamic strain with a switching frequency of 3.3 Hz. The corresponding BFS as a function of time are shown in (c) and (d). Step size of probe frequency scanning is 10 MHz.}
	\end{figure}

	We then proceed to switch the optical path, every 150 ms (switching frequency = 3.3 Hz). As explained earlier, a scan over 200 MHz with a step size of 3 MHz in the probe frequency requires a minimum of 210 ms. In order to resolve a dynamic strain at 3.3 Hz, we increase this step size to 10 MHz resulting in a BGS measurement time of 65 ms. The BGS traces and the corresponding BFS are shown in Fig. \ref{fig:75_80_BGS_150ms}.	
	The BGS traces obtained are as predicted and the BFS of the correlation peak location within Fiber 2 is in good agreement with the expected square wave fitting. The standard deviation in the estimated BFS is nearly 2 MHz.
	
\begin{figure}[htbp]
		\centering\subfloat[]{\includegraphics[width=6cm]{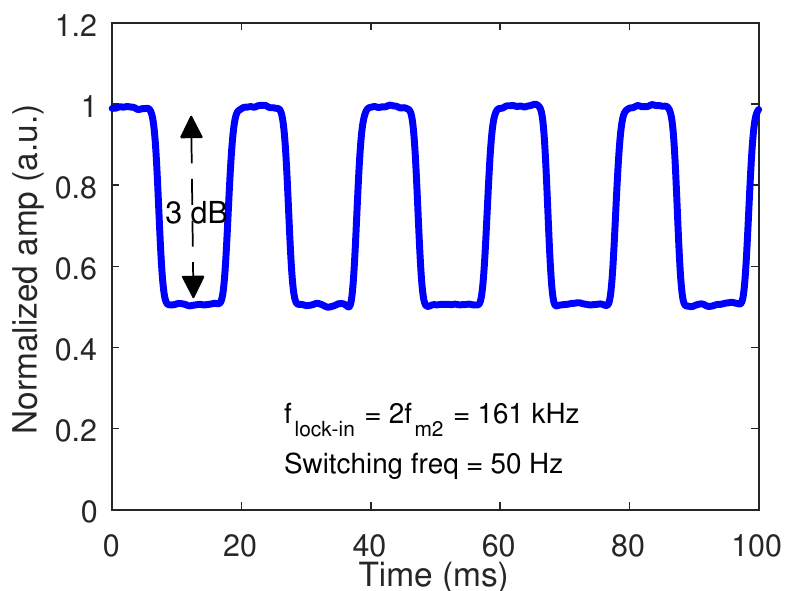}}
		\subfloat[]{\includegraphics[width=6cm]{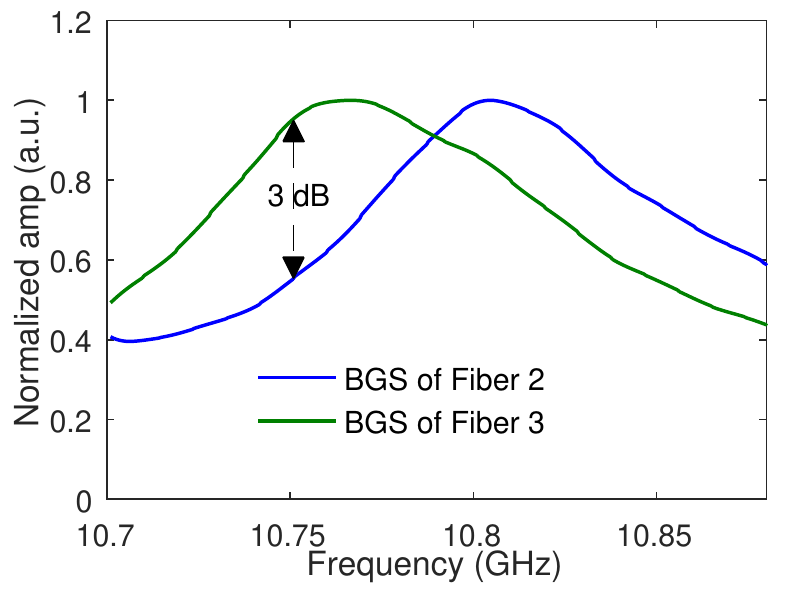}}
		\caption{\label{fig:80_10ms}(a) Amplified probe at a lock-in frequency of 161 kHz at a fixed pump-probe frequency offset of 10.750 GHz, when Fiber 2 is subjected to dynamic strain with a switching frequency of 50 Hz. BGS of the two fibers are shown in (b) for reference.}
	\end{figure}

	In the above experiments, the maximum rate of BFS variations that can be detected is limited by the finite sweeping time of the frequency offset between the pump and the probe. However, in situations where measurement of the absolute amplitude of strain is not very critical, perturbations at higher rates can be detected by measuring the intensity variations of the amplified probe at a specific pump-probe frequency offset. In order to demonstrate this, the optical path is switched between the two 100 m fibers every 10 ms (switching frequency = 50 Hz) and the frequency offset between the pump and the probe is fixed at 10.750 GHz. The amplified probe at a lock-in frequency of 161 kHz is monitored as a function of time and is shown in Fig. \ref{fig:80_10ms}(a).
	
	The amplitude variations shown in Fig. \ref{fig:80_10ms}(a) resemble a square wave with a frequency of 50 Hz which closely matches the frequency of dynamic strain variations. The extinction of the trace is nearly 3 dB, consistent with the change in Brillouin gain corresponding to the BGS shift as seen from the BGS traces in Fig. \ref{fig:80_10ms}(b). The rate of dynamic strain variations detected is limited by the switching speed of the optical switch used. Thus the phase modulation-based BOCDA technique is suitable to detect dynamic strain variations at multiple locations.
	
	In all the above experiments, an electrical spectrum analyzer in the zero span mode is used as a single channel lock-in amplifier where the two correlation peaks are monitored sequentially by only changing the lock-in frequency. This can be further extended to simultaneous monitoring of multiple locations by using a multi-channel lock-in amplifier, with each channel being locked to the corresponding 2$f_m$ frequency. For the proof-of-principle experiments demonstrated in this paper, the measurement range is 1.1 km and the spatial resolution is 6 m as decided by the choice of $f_m$ and $\Delta f$. The technique can further be extended to smaller FUT and sensing with better spatial resolution by choosing appropriate FM parameters. For instance, a measurement range of 10 m with a spatial resolution of 2 cm can be achieved with an $f_m$ in the range of 10 MHz and $\Delta f$ of 5 GHz.
	
	Another key aspect of this technique is that it is scalable and can be used to monitor the strain in multiple locations by generating multiple correlation peaks with appropriate sets of $f_m$ and $\Delta f$. In a typical BOCDA implementation, the highest measured frequency of dynamic strain is limited by the sampling rate of the receiver. In case of conventional BOCDA, simultaneous measurement of dynamic strain from two different sensing points would require a proportionately higher sampling rate. In contrast, the external phase modulation-based BOCDA provides a pathway to scale the number of sensing points while maintaining the original sampling rate, thereby preserving the maximum detectable frequency of dynamic strain at each sensing point. However, the scalability of sensing points would be constrained by the available RF and optical power levels shared among multiple FM signals which results in a compromise on the strength of the signal from each correlation peak. The resultant degradation in the signal-to-noise ratio (SNR) will have to be overcome by suitable post processing.
	
	\section{Conclusion}
	
	In this paper, we demonstrate a novel technique for multi-point sensing of dynamic strain variations using external phase modulation-based BOCDA which provides a clear pathway for monitoring multiple locations simultaneously. Multiple frequency modulations are generated with appropriate $f_m$ and $\Delta f$ values in the electrical domain, which are further transferred to the pump and the probe using external phase modulation. The BGS from multiple correlation peak locations are shown to be independent in the detection of strain variations through simulations, which are subsequently validated through controlled experiments. Two correlation peaks each 6 m wide are generated within the 1.1 km long FUT and the static strain variations at the two correlation peak locations are detected independently through lock-in detection at a frequency corresponding to twice the modulation frequency. We also demonstrate the detection of dynamic BFS variations at 3.3 Hz, only limited by the 65 ms sweep time of the probe frequency in our setup. By fixing the frequency offset between the pump and the probe, we demonstrate our capability to detect BFS variations at a rate of 50 Hz (limited by the speed of the optical switch used). Efforts are underway to scale the modulation frequency and the frequency deviation to achieve spatial resolution in the order of centimeters. We conclusively prove that the external phase modulation-based BOCDA technique is a viable solution for monitoring dynamic strain variations at multiple locations in a fiber simultaneously.
	
	\section*{Appendix}
	
	\subsection*{Generation of multiple correlation peaks}
	
	Consider the generation of two correlation peaks within the sensing fiber at two locations determined by the two modulation frequencies $f_{m1}$ and $f_{m2}$. The desired electric field of the pump at the input of the sensing fiber is given by
	
	\begin{eqnarray}
	E_{in}(t) = E_1 \exp\bigg[j\Big\{\big(\omega_c+\omega_1\big)t+\frac{\Delta f_1}{f_{m1}} \sin\big(2\pi f_{m1}t\big)\Big\}\bigg] \nonumber \\
	+E_2 \exp\bigg[j\Big\{\big(\omega_c+\omega_2\big)t+\frac{\Delta f_2}{f_{m2}} \sin\big(2\pi f_{m2}t\big)\Big\}\bigg] \label{E_reqd},
	\end{eqnarray}
	where $E_1$ and $E_2$ are the amplitudes, $\omega_c$ is the angular frequency of the optical carrier, $\omega_1$ and $\omega_2$ are the angular frequency difference between the optical carrier and center of two FM signals respectively, $f_{m1}$ and $f_{m2}$ are the modulation frequencies; and $\Delta f_1$ and $\Delta f_2$ are the frequency deviations of the two FM signals. The probe can be obtained by frequency shifting the pump by Brillouin frequency with an EOM in carrier suppressed configuration.
	
	Consider the case of generating the multiple FM signals in optical domain using the direct modulation of a light source. When the light source is modulated with a current consisting of sum of two sinusoids as shown below,
	\begin{eqnarray}
	I(t)=I_0+I_1 \cos\big(2\pi f_{m1}t\big)+I_2 \cos\big(2\pi f_{m2}t\big) \label{I(t)},
	\end{eqnarray}
	the corresponding electric field at the output of the laser is given by
	\begin{eqnarray}
	E_{laser}(t)=E_0 \exp\bigg[j\Big\{\omega_ct+\frac{\Delta f_1}{f_{m1}} \sin\big(2\pi f_{m1}t\big)+\frac{\Delta f_2}{f_{m2}} \sin\big(2\pi f_{m2}t\big)\Big\}\bigg], \label{E_laser}
	\end{eqnarray}
	which is not the same as in Eq. (\ref{E_reqd}). The electric field consists of only one carrier and hence cannot generate two FM signals simultaneously. Hence the direct modulation of a light source is not a viable solution in generating multiple independently-addressable correlation peaks.
	
	Now consider the case of generating the same using an external phase modulator. When the voltage signal driving the phase modulator consists of sum of two FM signals as shown below,
	\begin{eqnarray}
	V(t)=V_0\bigg[\sin\Big\{\omega_1t+\frac{\Delta f_1}{f_{m1}} \sin\big(2\pi f_{m1}t\big)\Big\}+\sin\Big\{\omega_2t+\frac{\Delta f_2}{f_{m2}} \sin\big(2\pi f_{m2}t\big)\Big\}\bigg] \label{V(t)}
	\end{eqnarray}
	the electric field at the output of the phase modulator is given by
	\begin{eqnarray}
	E_{mod}(t)=E_0 \exp\big(j\omega_ct\big) \exp\Big(j\frac{\pi V(t)}{V_\pi}\Big)
	\end{eqnarray}
	where $V_\pi$ is the half-wave voltage of the phase modulator. On expanding the second term using Taylor series, the electrical field is as below.
	\begin{eqnarray}
	E_{mod}(t)=E_0 \exp\big(j\omega_ct\big) \Big(1+j\frac{\pi V(t)}{V_\pi}+...\Big)
	\end{eqnarray}
	
	On substituting $V(t)$ with that in Eq. (\ref{V(t)}), expanding $\sin(\theta)$ and rearranging the terms, the electric field can be written as follows.
	\begin{eqnarray}
	E_{mod}(t)=E_0 \exp(j\omega_ct)+E_1\bigg[ \exp\bigg\{j\Big(\big(\omega_c+\omega_1\big)t+\frac{\Delta f_1}{f_{m1}} \sin\big(2\pi f_{m1}t\big)\Big)\bigg\} \nonumber \\
	+\exp\bigg\{j\Big(\big(\omega_c+\omega_2\big)t+\frac{\Delta f_2}{f_{m2}} \sin\big(2\pi f_{m2}t\big)\Big)\bigg\}\bigg] \nonumber \\
	-E_1\bigg[ \exp\bigg\{j\Big(\big(\omega_c-\omega_1\big)t-\frac{\Delta f_1}{f_{m1}} \sin\big(2\pi f_{m1}t\big)\Big)\bigg\} \nonumber \\
	+\exp\bigg\{j\Big(\big(\omega_c-\omega_2\big)t-\frac{\Delta f_2}{f_{m2}} \sin\big(2\pi f_{m2}t\big)\Big)\bigg\}\bigg]+...
	\end{eqnarray}
	
	If the higher order terms are filtered, then the electric field of the output of phase modulator consists of an optical carrier and multiple FM signals each at a center frequency of $\omega_c\pm\omega_1$ and $\omega_c\pm\omega_2$ with $f_m$ and $\Delta f$ decided by those corresponding to the voltage signals used to drive the phase modulator. Filtering higher frequencies can be achieved in the optical domain. Hence by using an external phase modulator, multiple FM signals in the electrical domain can be embedded in the optical domain and hence can be used to generate multiple independently-addressable correlation peaks.
	
	\section*{Funding}
	
	This work was supported by the Department of Electronics and Information Technology (DeitY), the Department of Science and Technology (DST), Office of the Principal Scientific Advisor, Government of India.
	
	\section*{Acknowledgments}
	
	Authors would like to thank Dr K V Reddy, PriTel Inc., USA for providing the optical amplifier.
	
\end{document}